# Nonlinear frequency conversion and manipulation of vector beams in a Sagnac loop

Chen Yang,[1, 2] Zhi-Yuan Zhou,[1, 2, 3,*] Yan Li,[1, 2] Yin-Hai Li,[1,2,3] Shi-Long Liu,[1, 2] Shi-Kai Liu,[1, 2] Zhao-Huai Xu,[1, 2] Guang-Can Guo,[1,2] and Bao-Sen Shi [1, 2,3,*]

[1] *CAS Key Laboratory of Quantum Information, USTC, Hefei, Anhui 230026, China*

[2] *Synergetic Innovation Center of Quantum Information & Quantum Physics, University of Science and Technology of China, Hefei, Anhui 230026, China*

[3] *Wang Da-Heng Collaborative Innovation Center for Science of Quantum Manipulation & Control, Heilongjiang Province & Harbin University of Science and Technology, Harbin 150080, China*

\* *Corresponding authors: zyzhouphy@ustc.edu.cn；drshi@ustc.edu.cn*

**Abstract:** Vector beams (VBs) are widely investigated for its special intensity and polarization distributions, which is useful for optical micromanipulation, optical micro-fabrication, optical communication, and single molecule imaging. To date, it is still a challenge to realize nonlinear frequency conversion (NFC) and manipulation of such VBs because of the polarization sensitivity in most of nonlinear processes. Here, we report an experimental realization of NFC and manipulation of VBs which can be used to expand the available frequency band. The main idea of our scheme is to introduce a Sagnac loop to solve the polarization dependence of NFC in nonlinear crystals. Furthermore, we find that a linearly polarized vector beam should be transformed to an exponential form before performing the NFC. The experimental results are well agree with our theoretical model. The present method is also applicable to other wave bands and second order nonlinear processes, and may also be generalized to the quantum regime for single photons.

## 1. INTRODUCTION

Laser technology has been developing for decades, most of which focus on spatially homogeneous states of polarization (SOP). Recently, laser beams with spatially inhomogeneous SOPs like vector-beam (VB) solutions of Maxwell's equations have attracted more and more attentions because of the unique intensity and polarization distribution in its transverse section [1]. One particular example with cylindrical symmetry of polarization is cylindrical vector beams (CVB) [2], including radial and azimuthal polarized vector beam, can be applied in high numerical aperture focusing [3], optical trapping [4], laser machining [5], optical cages [6-7], Super-resolution imaging [8], high capacity communication [9], and quantum information science [10-12]. Given the broad application prospects, researchers have presented several schemes to generate CVB, which can be classified into active [13, 14] and passive [15-17] schemes. In this letter, we use a passive scheme to generate CV beam, which is composed of two quarter wave plate (QWP), a half wave plate (HWP), and a vortex phase plate (VPP). A similar scheme was presented in [18], which was proved efficient and robust. Our scheme uses less optical elements and is more flexible.

As a very important laser technology, nonlinear frequency conversion (NFC) has been developed since the advent of lasers. It is a very important way to expand the available frequency range of CVB when it is difficult to generate it directly, such as the ultraviolet CVBs which is more useful in lithography technique because the shorter the wavelength is, the tighter the focus is. NFC of structured light beam have attracted a growing interest in the recent years [19-27]. Most NFC of structured light beam are focused on optical vortex beams with uniform polarization distributions [19-21]. The NFC of structured light beam with nonuniform polarization distributions was studied in theory [22-24] and experiment [25-27] only recently. Experimentally, extreme ultraviolet vector beams were created by driving infrared lasers through argon gas jet [26]; most recently, the polarization distribution of full Poincaré beams in second harmonics generation (SHG) was studied by using a type-II phase-matching KTP crystal [27]. To date, it is still a challenge to realize NFC of CVBs due to the polarization sensitive phase matching condition in NFC.

In this letter, we demonstrate a scheme to realize SHG of a CVB with only one nonlinear crystal, which is also applicable to other second order nonlinear processes such as sum-frequency generation, difference frequency generation, and may also be generalized to the quantum regime for single photons. The main idea of our scheme is using a Sagnac loop to solve the polarization sensitivity in NFC. Another important feature of method is that in order to generate arbitrary CVBs in the SHG band, the pump beam CVBs should be transformed into the exponential form. We first build a general theoretical model for the generation and NFC of CVB, then we verify the theoretical model experimentally. CVBs with different topological charges are frequency converted and manipulated in our experiment. Arbitrary CVBs with different polarization distributions can be can be obtained. The experimental results are in full agreement with our theoretical model.

We would like to introduce the general theoretical model at first, and then show the experimental demonstration. The Fig.1 shows the experimental setup, and also is a theory schematic.

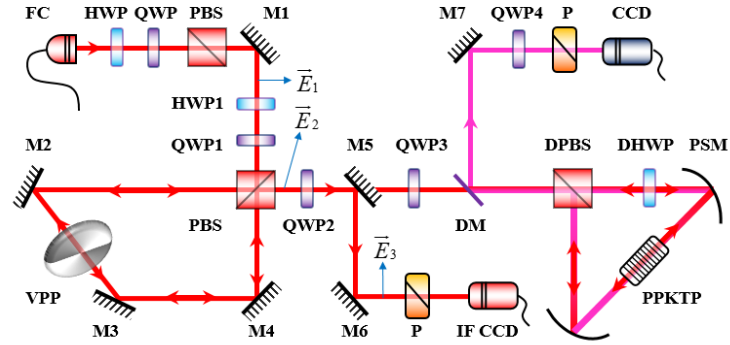

Fig.1. Schematic of the experimental setup. FC, fiber collimator; HWP, half wave plate; QWP, quarter wave plate; PBS, polarization beam splitter; M, mirror; VPP, vortex phase plate; DM, dichroic mirror; P, polarizer; DPBS, dichroic PBS; DHWP, dichroic HWP; PSM, parabolic silver mirror; PPKTP, periodically poled potassium titanyl phosphate crystal; IF CCD, infrared charge coupled device.

## 2. THEORY

The pump CVB at fundamental wavelength of 1560nm is generated from a Sagnac loop with a vortex phase plate (VPP). The mathematical expression of the target linearly polarized CVB is $\left[\vec{e}_x \cos(l\varphi+\theta) + \vec{e}_y \sin(l\varphi+\theta)\right] A(\mathrm{r})$ where $(\varphi, r)$, $l$, $\theta$ is the space coordinate, the topological charge, the initial phase of the CVB. $A(\mathrm{r})$ is the space dependent amplitude term, which will be omitted in the following calculations because it's independent of the space azimuthal angle and is the common term of the two orthogonal polarized parts. We use a cascading system to generate this beam. The operation principle of the generation of CVB for the pump are as follows: Firstly, let a horizontal polarized light $\vec{E}_1 = \begin{pmatrix} 1 \\ 0 \end{pmatrix}$ pass through HWP1 and QWP1 with their fast axis forming an angle of $\alpha$ and $-\pi/4$ with the horizontal direction respectively. Next, a Sagnac loop with a VPP is used to generate the hybrid superposition modes with polarization and orbital angular momentum (OAM), in which a polarization beam splitter (PBS) separates horizontal and vertical polarization. The Jones matrix of the Sagnac loop can be written as

$$S(l) = \begin{pmatrix} e^{il\varphi} & 0 \\ 0 & e^{-i(l\varphi+\Delta_1)} \end{pmatrix}.$$ 

Here $\Delta_1$ is the phase difference in Sagnac loop which is caused by reflection of mirrors and birefringence of PBS. The polarization and OAM superposition mode can be calculated as

$$\vec{E_2} = S(l) Q\left(-\frac{\pi}{4}\right) H(\alpha) \vec{E_1} = \frac{1}{\sqrt{2}} \begin{pmatrix} e^{i(l\varphi+2\alpha)} \\ ie^{-i(l\varphi+2\alpha+\Delta_1)} \end{pmatrix}, \quad (1)$$

where $Q$ and $H$ stand for the Jones matrix of HWP and QWP respectively. This beam is a hybrid polarized CVB [28], which can be transformed into linearly polarized vector beam by only QWP2 with its fast axis orienting at $\pi/4$. The linearly vector beam can be described as

$$\vec{E_3} = Q\left(\frac{\pi}{4}\right) \vec{E_2} = e^{-i\Delta_1/2} \begin{pmatrix} \cos(l\varphi+\theta) \\ \sin(l\varphi+\theta) \end{pmatrix} \quad (2)$$

Where $\theta = 2(\alpha + \Delta_1/4)$ is the initial phase of the CVB which can be controlled by adjusting the optical axis angle $\alpha$ of HWP1. The schematic is shown in the left half of Fig.1 before M5. And then the right half shows the SHG of CVB based on the second Sagnac loop.

In the second Sagnac loop, a dichroic PBS (DPBS) is used to separate the horizontal and vertical polarization, one of which is transformed to the orthogonal polarization by a dichroic HWP (DHWP) before the crystal. So the fundamental beam pump the nonlinear crystal from opposite directions but with the same polarization. Then the two SH beams in same polarization are generated, before the DPBS, one of the SH beam is rotated to its orthogonal polarization by the DHWP, therefore the two SH beams are combined at the DPBS in orthogonal polarizations. In our theory and experiment we take a type-0 (zzz) crystal as example, which only responds to vertical polarization. So the DHWP is set in the transmission side as shown in Fig.1, and the counter-propagating beams will return along the original path after combination. And then the pump beam passes through the dichroic mirror (DM) and the SH beam is separated by reflection.

The transformation of the polarization states in the Sagnac can be described by an operator

$$T = \begin{pmatrix} 0 & 1 \\ e^{i\Delta_2} & 0 \end{pmatrix},$$

where $\Delta_2$ is the phase difference in the second Sagnac loop mainly caused by the asymmetric position of the crystal. Besides the linear process, the nonlinear process can't be represented by Jones matrixes. Considering the relation $E^{2\omega} \propto (E^{\omega})^2$ between the SH beam and the fundamental beam under paraxial and undepleted-pump approximation, the complete process in the Sagnac loop could be represented as $E^{2\omega} \propto T(E^{\omega})^2$.

It's remarkable that the linearly polarized vector beam $\vec{E_3}$ can't be frequency doubled directly, or it will generate accompanying Gauss beam component in the process, which can be explained by the following equation,

$$\vec{E}_3^{2\omega} \propto T(\vec{E}_3)^2 = T\begin{pmatrix} \cos^2(l\varphi+\theta) \\ \sin^2(l\varphi+\theta) \end{pmatrix} = \frac{1}{2}\begin{pmatrix} 1-\cos(2l\varphi+2\theta) \\ 1+\cos(2l\varphi+2\theta) \end{pmatrix} \quad (3)$$

The constant term stands for the Gauss beam and another term is linearly polarized superposed OAM beams, which is not the expected result. While the square of exponential form does not produce new terms. So if we need to frequency doubling a CVB, it has to be transformed into a hybrid polarized CVB like $\vec{E}_2$ by QWP3 at first (The fast axis angle of QWP3 is opposite to that of QWP2, which means we obtain $\vec{E}_2$ again). After frequency doubling of $\vec{E}_2$, we use QWP4 with its fast axis orienting at $\pi/4$ to obtain linearly polarized vector beam. The process can be described as

$$\vec{E}_2^{2\omega} \propto Q\left(-\frac{\pi}{4}\right)T(\vec{E}_2)^2 = \begin{pmatrix} \cos(2l\varphi+\Theta) \\ \sin(2l\varphi+\Theta) \end{pmatrix} e^{-i(\Delta_1-\Delta_2/2-3\pi/4)} \quad (4)$$

where $\Delta_2$ is the phase difference in the second Sagnac loop, $\Theta = 4\alpha + \Delta_1 + \Delta_2/2 - \pi/4$ is the initial phase of the CVB, which is still be controlled by HWP1. Neglecting the global phase, we have obtained the second harmonic CVB in theory.

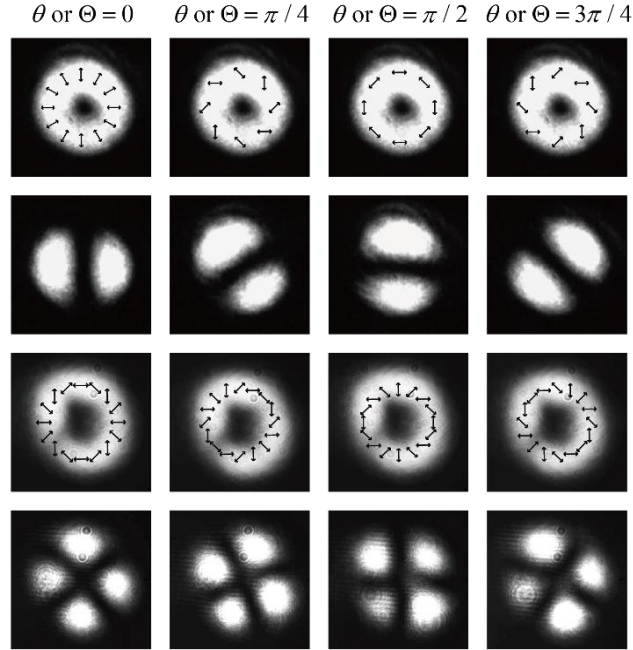

Fig.2 Fundamental (the top and the second rows) and SH CVBs (the third and the bottom rows) with $l=1$; the arrows indicate the SOP distribution, and the second and the bottom rows show the results measured by horizontal polarizer.

### 3. EXPERIMENTAL DEMONSTRATION

In experiment, the 1550 nm continuous-wave pump beam is derived from a diode laser (Toptica prodesign) and is amplified by an erbium-doped fiber amplifier. A PBS is used to obtain horizontal polarization beam

$\vec{E}_1$. A QWP and a HWP before the PBS are used to control amplitude of the pump beams, while the HWP1 and QWP2 control the phase difference. Due to the QWP1 with fast axis orientating at $-\pi/4$, a circular polarization is generated before Sagnac loop, which makes sure the two counter-propagating orthogonal polarized beams in the Sagnac loop have equal energy. In order to ensure the same spatial propagation and evolution of the two counter-propagating beams, we place the VPP at the middle of the Sagnac loop. Considering the topological charge will be opposite after reflection, every beam with $l \neq 0$ should be reflected by even mirrors, so the Sagnac loop is designed as a quadrangle. According to the theory, a linearly polarized CVB will be generated after QWP2, so we use mirrors M5 and M6 to separate it. And we record the intensity distribution with an infrared charge coupled device (CCD). The results with $l=1$ is shown in the top row of Fig.2. The arrows stand for the polarization direction. The hole in the center of the field arises from the phase singularity. In order to check the polarization distribution, we use a polarizer to perform projection measurement. The second row of Fig.2 shows the results of measurement with a horizontal direction polarizer. Only the polarization in the same direction as the polarizer can be recorded, while the orthogonal polarization is extinct.

After recording the fundamental CVB, we remove the mirror M5 to perform the SHG experiment. The pump laser is focused on the center of a type-0 periodically poled potassium titanyl phosphate (PPKTP) crystal by two symmetric parabolic silver mirrors with focal length of 101.6mm. The PPKTP crystal has dimension of 1 mm $\times 2$ mm $\times 8$ mm, and is fixed at the middle of Sagnac loop. A homemade temperature controller controls the temperature of the PPKTP crystal with a stability of 2mK. A DM is used to separate the SHG beam, which is measured by a polarizer and finally recorded by a visible band CCD.

The third and the bottom row of Fig.2 show the SHG experiment results with their fundamental beam corresponding the same columns of the top row. As shown in equation (4), the topological charge of the second harmonic CVB has been doubled due to the conservation of angular momentum [21]. So the results of four petals in Fig.2 are consistent with the theory. Fig.3 shows the experiment results of fundamental CVBs and their second harmonic beams with $l=2,3,4$ and $\theta=0$, which also support the theory. Actually, for a CVB with topological charge $l$, every polarization occurs $2l$ times within a complete circle, so there will be $2l$ petals no matter what direction of the polarizer. In our experiment, the frequency conversion efficiency is just about 0.01% when the pump power is 500mW. It can be improved by using a strong pulse pump or inserting a symmetric confocal resonator in the Sagnac loop in future experiments.

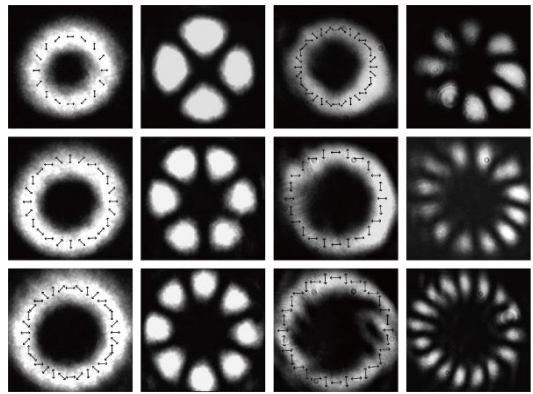

Fig.3. Fundamental (the first and the second columns) and SH CVBs (the third and the fourth columns) with $l = 2,3,4$ (the top row to the bottom row); the arrows indicate the SOP distribution, and the second and the fourth columns show the results measured by horizontal polarizer.

## 4. DISCUSSION AND CONCLUSION

According to the relations $\theta=2(\alpha+\Delta_1/4)$ and $\Theta = 4\alpha + \Delta_1 + \Delta_2/2 - \pi/4$, it is convenient to control $\theta$ and $\Theta$ by tuning $\Delta_1$ and $\Delta_2$ through rotating only HWP1 in experiment by using this configuration.

In addition, if we use a HWP and a QWP instead of the only QWP2 after the first loop, the hybrid polarized vector beam can be transformed into arbitrary hybrid polarized CVB, which provides the additional degree to control the spatial structure of polarization and to engineer the focusing field [29]. As we know, a circular polarization can be transformed to an arbitrary polarization by a HWP and a QWP. Moreover, a hybrid polarized CVB is the superposition of two orthogonal elliptically polarized beams with OAM [29,30], and a linearly polarized CVB is the superposition of two orthogonal circular polarized beams with OAM. The operation can also works to the second harmonic CVB if we replace QWP4 with a pair of wave plates. So we could obtain arbitrary fundamental or SHG hybrid polarized CVB by using this configuration.

More generally and interestingly, if we rotate the QWP1 which means changing the energy proportion of the two counter-propagating orthogonal polarized beams in the Sagnac loop, we will generate an arbitrary CV beam whose polarization is distributed on a circle of arbitrary size and position on the Poincaré sphere with a new degree of freedom. And all the states on high-order Poincaré sphere [30] can be generated by using this configuration.

In the future, the present scheme can be modified with much more flexibility if we replace VPP with a spatial light modulator [31, 32], by which we can study the NFC of CVB with fractional topological charges. Furthermore, we can generalize the present scheme to other second order nonlinear processes. Specially, the sum-frequency generation enable us select target frequency band and topological charge with more flexibility.

In conclusion, we realize nonlinear frequency conversion of CV beams based on a Sagnac loop. And we find that the CV beam should be transformed to exponential form before performing the NFC and transformed back after NFC. Furthermore, based on this configuration, an arbitrary hybrid polarized CVB can be generated. The method and theoretical analysis we presented are also applicable to other wave bands and second order nonlinear processes, and may also be generalized to the quantum regime for single photons.

**Funding**

Natural Science Foundation of China (NSFC) (61605194, 61435011, 61525504); Anhui Initiative In Quantum Information Technologies (AHY020200); the China Postdoctoral Science Foundation (2017M622003); the National Key Research and Development Program of China (2016YFA0302600).